\documentclass[12pt]{article}
\usepackage{graphicx}
\usepackage{amssymb}
\newlength{\extraspace}
\newlength{\extraspaces}
\newcommand{\be}{\begin{equation}
\addtolength{\abovedisplayskip}{\extraspaces}
\addtolength{\belowdisplayskip}{\extraspaces}
\addtolength{\abovedisplayshortskip}{\extraspace}
\addtolength{\belowdisplayshortskip}{\extraspace}}
\newcommand{\ee}{\end{equation}}

\newcommand{\ba}{\begin{eqnarray}
\addtolength{\abovedisplayskip}{\extraspaces}
\addtolength{\belowdisplayskip}{\extraspaces}
\addtolength{\abovedisplayshortskip}{\extraspace}
\addtolength{\belowdisplayshortskip}{\extraspace}}
\newcommand{\ea}{\end{eqnarray}}

\newcounter{saveeqn}

\newcommand{\dif}{\mathrm{d}}

\begin{document}
\addtolength{\baselineskip}{1.5mm}

\thispagestyle{empty}
\begin{flushright}

\end{flushright}
\vbox{}
\vspace{2cm}

\begin{center}
{\LARGE{A new AF gravitational instanton%\\[2mm]
        }}\\[16mm]
{{Yu Chen}~~and~~{Edward Teo}}
\\[6mm]
{\it Department of Physics,
National University of Singapore, %\\[1mm]
Singapore 119260}\\[15mm]

\end{center}
\vspace{2cm}

\centerline{\bf Abstract}
\bigskip
\noindent
It has long been conjectured that the Euclidean Schwarzschild and Euclidean Kerr instantons are the only non-trivial asymptotically flat (AF) gravitational instantons. In this letter, we show that this conjecture is false by explicitly constructing a new two-parameter AF gravitational instanton with a $U(1)\times U(1)$ isometry group, using the inverse-scattering method. It has Euler number $\chi=3$ and Hirzebruch signature $\tau=1$, and its global topology is $\mathbb{C}P^2$ with a circle $S^1$ removed appropriately. Various other properties of this gravitational instanton are also discussed.

%\addtolength{\baselineskip}{3mm}   %double-spacing

\newpage

Gravitational instantons are the analogues of Yang--Mills instantons in Einstein's general relativity. In Euclidean quantum gravity \cite{Gibbons:1994}, they are stationary phase points in the path integral for the amplitude to tunnel between two field configurations. So they are classical solutions to Einstein's field equations, and are required to be non-singular and have positive-definite signature. When the cosmological constant vanishes, which will be the case considered in this letter, they are nothing but Ricci-flat Riemannian 4-manifolds. In this case, all the explicitly known gravitational instantons \cite{Gibbons:1976ue,Hawking:1976jb,Eguchi:1978xp,Page:1979aj,Gibbons:1979zt}, except $T^4$ (with flat metric), are non-compact with ``infinities''.

There is an interesting class of gravitational instantons which are said to be asymptotically flat (AF) \cite{Gibbons:1979gd,Chen:2010zu}. This means that outside some compact subset, their underlying manifolds are diffeomorphic to $S^1\times \mathbb{R}\times S^2$, and near infinity $r\rightarrow\infty$, their metrics approach the form
\be
\label{definition_AF}
\dif s^2\rightarrow \dif \psi^2+\dif r^2+r^2\left(\dif\theta^2+\sin^2 \theta\,\dif\phi^2\right ),
\ee
at least as fast as $r^{-1}$, with identifications made on $\psi$ (possibly together with a translation along $\phi$).  Such gravitational instantons contribute to the thermal canonical ensemble of gravitational fields \cite{Gibbons:1976ue}. The flat space $S^1\times \mathbb{R}^3$ is trivially an AF gravitational instanton, but it is not the only one. Other known AF gravitational instantons are the Euclidean Schwarzschild and Euclidean Kerr instantons \cite{Gibbons:1976ue}.  This is in contrast to asymptotically Euclidean (AE) gravitational instantons \cite{Gibbons:1979gd}, which asymptote to standard Euclidean space at infinity. It has been shown that the flat space $E^4$ is in fact the unique AE gravitational instanton \cite{Schon:1979uj}.

Now, it has been conjectured that the Euclidean Schwarzschild and Euclidean Kerr instantons are the only AF gravitational instantons other than flat space $S^1\times \mathbb{R}^3$ (see, e.g., \cite{Gibbons:1979xm,Lapedes:1980st}). This conjecture was originally made on the basis that the classical black-hole uniqueness theorems should apply to the Euclidean regime. However, Lapedes \cite{Lapedes:1980st} has pointed out that although a Euclidean version of Israel's theorem provides a type of uniqueness theorem for the Euclidean Schwarzschild instanton, a Euclidean version of Robinson's theorem does not allow one to form conclusions about the uniqueness of the Euclidean Kerr instanton. It has remained an unsolved problem to prove or disprove the above-mentioned conjecture in Riemannian geometry for the past three decades.

Recently, the present authors classified gravitational instantons with an isometry group $\mathcal{T}\equiv U(1)\times U(1)$ using the rod-structure formalism \cite{Chen:2010zu}, and found that there exists a new rod structure, Fig.~5(a) in \cite{Chen:2010zu} or Fig.~\ref{Figure_new_AF} below, that may be associated with some as yet undiscovered AF gravitational instanton. We refer the reader to \cite{Chen:2010zu} and references therein for more details on the rod-structure formalism; here we only give a brief summary of its basic idea. The rod structure assigns a sequence of so-called rods or intervals along a certain $z$ axis, $(-\infty,z_1]$, $[z_1,z_2]$,\dots, $[z_{N-1},z_N]$, $[z_N,\infty)$, to a given gravitational instanton with an isometry group $\mathcal{T}$, with each rod having a certain direction. The direction of a rod is a normalised Killing vector field which vanishes along that rod; it generates a $U(1)$ isometry subgroup of $\mathcal{T}$, and has (principal) orbits with period $2\pi$. The points where adjacent rods meet are called turning points. We can identify any direction pair of adjacent rods as the pair of independent $2\pi$-periodic generators of the isometry group $\mathcal{T}=U(1)\times U(1)$ of the gravitational instanton, and all such direction pairs of adjacent rods must be related by $GL(2,\mathbb{Z})$ transformations. Each rod in the rod structure represents a two-dimensional fixed-point set, a ``bolt'' in the terminology of \cite{Gibbons:1979xm}, of the $U(1)$ isometry subgroup generated by its direction. On the other hand, a turning point represents an isolated fixed point of the whole isometry group $\mathcal{T}$; it is a ``nut'' \cite{Gibbons:1979xm} with respect to any Killing vector field which generates isometries in $\mathcal{T}$, provided the Killing vector field is linearly independent with each of the directions of the two rods meeting at that turning point.

In this letter, we explicitly construct a new AF gravitational instanton with the rod structure shown in Fig.~\ref{Figure_new_AF}. This gravitational instanton provides a counterexample to the conjecture described above (in particular, it is a counterexample to Conjecture II as precisely formulated by Lapedes \cite{Lapedes:1980st}).  Its existence also implies that the classical black-hole uniqueness theorems can in no way be taken over to the Euclidean regime in general.\footnote{However, Simon \cite{Simon:1995ty} has proved the uniqueness of the Euclidean Kerr instanton under additional assumptions.}

The metric of the new AF gravitational instanton we obtained can be written in C-metric-like coordinates in the following form:
\be
\label{metric_new_AF}
\dif s^2={\frac {F(x,y) }{H(x,y) }} \left( {\dif \psi}+\Omega
 \right) ^{2}-\frac{{\varkappa}^{4} H(x,y) }{\left( x-y \right) ^{3}}\left[ {\frac {{{\dif x}}^{2}}{G(x) }}-{\frac
{{{\dif y}}^{2}}{G(y) }}+\frac{4G(x)G(y)}{(x-y)F(x,y)}\,{{\dif \phi}}^{2} \right],
\ee
where the one-form $\Omega$, and the functions $G(x)$, $H(x,y)$ and $F(x,y)$ are defined as
\ba
\Omega&=&{\frac {{\varkappa}^{2} \left( 1-{\lambda}^{2} \right)  \left( 1+\gamma
 \right)  \left( x+\lambda \right)  \left( y+\lambda \right)   }{ 2\left( 2+\gamma \right)  \left( x-y
 \right) F(x,y) }}\Big\{2\left( x+1
 \right)  \left( y-1 \right)\cr&&
 \times\left[ \lambda\left( 3+\gamma
 \right)\left(2  \left( x+\lambda \right)  \left( y+\lambda \right) +\left( 1+\gamma \right)  \left( \gamma x-\gamma y
-2 \right)\right) +3\left( 1-{\lambda}^{2} \right)  \left( x+y
 \right)  \right] \cr&&
  +\left( 1+\lambda \right) ^{3} \left( \gamma-\lambda+2 \right) ^{2}
 \left( {x}^{2}-1 \right) - \left(1- \lambda \right) ^{3} \left(
\gamma+\lambda+2 \right) ^{2} \left( {y}^{2}-1 \right)\Big\}\,
\dif \phi\,,\cr
G(x)&=&(1-x^2)(x+\lambda)\,,\cr
H(x,y)&=&{\frac { \left( \lambda+\gamma \right)  \left( x+\lambda
 y \right) + \left( \lambda-\gamma \right)  \left( y-\lambda x
 \right) +2xy+2\gamma{\lambda}^{2}  }{4(2+\gamma)
}} \Big\{\left( 2+\gamma \right)  \left( x+\lambda \right)
  \cr
&&\times\left( y+\lambda \right) \left[ \lambda\left( 1+\gamma \right)
 \left( x+y \right) + \left( {\lambda}^{2}-\gamma \right)  \left( x-y
 \right) +4+2\gamma{\lambda}^{2}-2xy \right] \cr&&
 + \left( 1-{\lambda}^{2} \right)
\left[  \left( \gamma+\lambda+2 \right)\left(
x+\lambda \right)  \left( x+\lambda\gamma \right) + \left( \gamma-
\lambda+2 \right)  \left( y+\lambda \right)  \left( y+\lambda\gamma
 \right)  \right]  \Big\}\,,\cr
F(x,y)&=&\left( x+\lambda \right)  \left( y+\lambda \right)  \big[  \left( 1+\lambda\gamma \right) ^{2} \left( xy+ \lambda x+\lambda y +1
 \right) -2\lambda\gamma\left( 1-\lambda \right)  \left( x-1
 \right)  \left( y-1 \right) \cr
 &&- \left( {x}^{2}-1 \right)  \left( {y}^{2
}-1 \right)  \big]\,,
\ea
with $\gamma\equiv\sqrt{2-\lambda^2}$. The parameters $\varkappa$,  $\lambda$ and coordinates $x$, $y$ take the ranges $0<\varkappa<\infty$, $-1<\lambda<1$, $-\infty\leq x \leq -1$ and $1\leq y \leq \infty$.

The above solution admits an obvious isometry group $\mathcal{T}=U(1)\times U(1)$ generated by the two Killing vector fields $\frac{\partial}{\partial \psi}$ and $\frac{\partial}{\partial \phi}$. In what follows, we analyse the rod structure of this solution, closely following \cite{Chen:2010zu}. The Weyl--Papapetrou coordinates $(\psi,\phi,\rho,z)$ are related to the above coordinates by
\be
\label{Weyl_coordinates}
\rho={\frac {2{\varkappa}^{2}\sqrt {-G ( x ) G ( y ) }}{
 \left( x-y \right) ^{2}}}\,, \qquad z={\frac {{\varkappa}^{2} \left(1-xy \right)  \left( x+y+2\lambda \right) }{
 \left( x-y \right) ^{2}}}\,.
\ee
In these coordinates, the rod structure has three turning points, at $(\rho=0,z=z_1\equiv-\varkappa^2)$ or $(x=-\infty,y=1)$, $(\rho=0,z=z_2\equiv \lambda \varkappa^2)$ or $(x=-1,y=1)$,  and $(\rho=0,z=z_3\equiv \varkappa^2)$ or $(x=-1,y=\infty)$. They divide the $z$-axis into four rods; from left to right they are:

\begin{itemize}
\item Rod 1: a semi-infinite rod located at $(\rho=0, z\leq z_1)$ or $(x=-\infty,1\leq  y< \infty)$, with direction $\ell_1=(0,1)$.

\item Rod 2: a finite rod located at $(\rho=0, z_1\leq z\leq z_2)$ or $(-\infty\leq x\leq -1,y=1)$, with direction
\be
\ell_2=\left({\frac {{\varkappa}^{2} \left( 1-{\lambda}^{2} \right)^2  \left( 1+\gamma
 \right) ^{2}}{2\left(\gamma-\lambda^2\right)}}
,-\frac{\left( \lambda+\gamma \right) ^{2}}{4}\right ).
\ee
\item Rod 3: a finite rod located at $(\rho=0, z_2\leq z\leq z_3)$ or $(x=-1,1\leq y\leq \infty)$, with direction
\be
\ell_3=\left({\frac {{\varkappa}^{2} \left( 1-{\lambda}^{2} \right)^2  \left( 1+\gamma
 \right) ^{2}}{2\left(\gamma-\lambda^2\right)}}
,\frac{\left( \lambda-\gamma \right) ^{2}}{4}\right ).
\ee

\item Rod 4: a semi-infinite rod located at $(\rho=0, z\geq z_3)$ or $(-\infty<x\leq -1,y=\infty)$, with direction $\ell_4=(0,1)$.
\end{itemize}
Note that when $\lambda=0$, rods 2 and 3 have the same length.

To simplify the expressions for future purposes, we can formally write $\ell_2=\left(\frac{1}{\kappa_E},\frac{\Omega_{1E}}{\kappa_E}\right)$ and $\ell_3=\left(\frac{1}{\kappa_E},\frac{\Omega_{2E}}{\kappa_E}\right)$, where
\be
\label{kappa_omega}
\kappa_E={\frac {2(\gamma-\lambda^2)}{{\varkappa}^{2} \left( 1-{\lambda}^{2} \right)^2  \left( 1+
\gamma \right) ^{2}}}\,,\quad \Omega_{1E}=-{\frac {2(\gamma-{\lambda}^{2})}{{\varkappa}^{2} \left( \gamma-\lambda
 \right) ^{2} \left( 1+\gamma \right) ^{2}}}\,,\quad \Omega_{2E}={\frac {2\left(\gamma-{\lambda}^{2}\right)}{{\varkappa}^{2} \left( \lambda+\gamma
 \right) ^{2} \left( 1+\gamma \right) ^{2}}}\,.
\ee
We can check that $\ell_3=\ell_1+\ell_2$ as required. To ensure regularity, the orbits generated by any direction pair of adjacent rods, say $\{\ell_1,\ell_2\}$, should be identified with period $2\pi$ independently, i.e.,
\be
\label{identifications}
(\psi, \phi) \rightarrow (\psi, \phi+2 \pi) \,,\qquad (\psi, \phi) \rightarrow \left(\psi+\frac{2\pi}{\kappa_E}, \phi+\frac{2\pi \Omega_{1E}}{\kappa_E}\right) .
\ee
The direction pair $\{\ell_1,\ell_2\}$ is then identified as the pair of independent $2\pi$-periodic generators of the isometry group $\mathcal{T}$ of this gravitational instanton.

We can put the rod structure of this gravitational instanton in standard orientation \cite{Chen:2010zu} by introducing new Weyl--Papapetrou coordinates $(\tilde{\psi},\tilde{\phi},\tilde{\rho},\tilde{z})$, which are related to the old ones (\ref{Weyl_coordinates}) by
\be
\psi=\frac{1}{\kappa_E}\,\tilde{\psi}\,,\qquad
\phi=\frac{\Omega_{1E}}{\kappa_E}\,\tilde{\psi}+\tilde{\phi}\,,\qquad
\rho=\kappa_E \tilde{\rho}\,,\qquad
z=\kappa_E \tilde{z}\,.
\ee
These new coordinates are chosen such that $\{ \frac{\partial}{\partial \tilde{\psi}}=\ell_2, \frac{\partial}{\partial \tilde{\phi}}=\ell_1 \}$, so in the new coordinate basis $\{\frac{\partial}{\partial \tilde{\psi}},\frac{\partial}{\partial \tilde{\phi}}\}$, the four rod directions now take very simple forms, i.e., $K_1=(0,1)$, $K_2=(1,0)$, $K_3=(1,1)$, and $K_4=(0,1)$. The three turning points are now pushed to $(\tilde{\rho}=0,\tilde{z}=\tilde{z_i}\equiv \frac{z_i}{\kappa_E})$ respectively. This is depicted in Fig.~\ref{Figure_new_AF}. In the new Weyl--Papapetrou coordinates, the following identifications should be made to ensure regularity:
\be
(\tilde{\psi}, \tilde{\phi}) \rightarrow (\tilde{\psi},\tilde{ \phi}+2 \pi) \,,\qquad (\tilde{\psi}, \tilde{\phi}) \rightarrow (\tilde{\psi}+2 \pi, \tilde{\phi}) \,.
\ee

\begin{figure}[t]
\begin{center}
\includegraphics{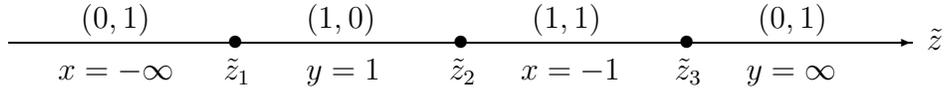}
\caption{The rod structure of the new AF gravitational instanton in standard orientation.}
\label{Figure_new_AF}
\end{center}
\end{figure}

The curvature invariants of the gravitational instanton (\ref{metric_new_AF}) are finite, and its signature is positive-definite everywhere. This gravitational instanton, as a 4-manifold with a natural $\mathcal{T}=U(1)\times U(1)$ action (as an isometry), is uniquely determined by its rod structure and can be read off from it \cite{Hollands:2008fm,Chen:2010zu}. Note that in Fig.~\ref{Figure_new_AF} the directions of rods 1 and 4 are identical. If we assume that these two rods are joined up to a single rod at infinity, we obtain the rod structure of $\mathbb{C}P^2$ \cite{Chen:2010zu}. By this assumption, we are actually adding a manifold $S^1\times \mathbb{R}^3$ to the gravitational instanton. So the topology of (\ref{metric_new_AF}) is in fact $\mathbb{C}P^2-S^1\times\mathbb{R}^3\cong \mathbb{C}P^2-S^1$.\footnote{A similar argument applied to the Euclidean Kerr instanton would show that its topology is $S^4-S^1\cong\mathbb{R}^2\times S^2$, in accordance with known results \cite{Gibbons:1979xm}.} Note that $\mathcal{T}$ acts on the removed part $S^1\times \mathbb{R}^3$ in the usual way, i.e., one of the $U(1)$'s generates the $S^1$, while the other generates the rotational symmetry of $\mathbb{R}^3$. The common boundary is then $S^1\times S^2$ with the obvious $\mathcal{T}$ action.

The two semi-infinite rods $x=-\infty$ (rod 1) and $y=\infty$ (rod 4) represent the two rotational axes of $\frac{\partial}{\partial \phi}$ that extend to infinity, while the two finite rods $x=-1$ (rod 3) and $y=1$ (rod 2) represent two non-contractible $S^2$ surfaces. The areas of these two $S^2$ surfaces are $\frac{2\pi \varkappa^2}{\kappa_E}(1-\lambda)$ and $\frac{2\pi \varkappa^2}{\kappa_E}(1+\lambda)$ respectively. On the other hand, (hyper-)surfaces of constant $x$ for $-\infty<x<-1$ touch rods 2 and 4, so they have topology $S^3$. They enclose the manifold of a Taub-bolt instanton \cite{Page:1979aj}, i.e., $\mathbb{C}P^2$ with a (nut) point removed. This is confirmed by the fact that we recover the Taub-bolt instanton by pushing $\tilde z_1$ to $-\infty$ while keeping rod 3 finite. A similar situation happens for surfaces of constant $y$ for $1<y<\infty$, and we again recover the Taub-bolt instanton by pushing $\tilde z_3$ to $\infty$ while keeping rod 2 finite. The two manifolds enclosed by constant $-\infty< x< -1$ and $1<y<\infty$ surfaces intersect in a single nut region $\mathbb{R}^4$. We mention that in the limits $\lambda=\pm 1$, we recover a three-dimensional flat space.

Physical infinity of the gravitational instanton is located at $(x\rightarrow -\infty,y\rightarrow \infty)$, where the metric (\ref{metric_new_AF}) approaches (\ref{definition_AF}) for $r\rightarrow \infty$ if we perform the coordinate transformations $\left\{x=-\frac{r}{\varkappa^2}\csc^2\frac{\theta}{2},y=\frac{r}{\varkappa^2}\sec^2\frac{\theta}{2}\right\}$. Together with the identifications (\ref{identifications}), we can see that this gravitational instanton is indeed AF. Its asymptotic geometry is exactly the same as that of the Euclidean Kerr instanton \cite{Chen:2010zu}, if the Euclidean surface gravity and angular velocity of the latter are respectively identified with $\kappa_E$ and $\Omega_{1E}$ in (\ref{kappa_omega}). In particular, as mentioned above, infinity of this gravitational instanton has topology $S^1\times S^2$. The Killing vector field $\frac{\partial}{\partial \tilde{\psi}}$, whose orbits close off along rod 2, generates the $S^1$ (which in fact become infinitely large), while $\frac{\partial}{\partial \tilde{\phi}}=\frac{\partial}{\partial \phi}$ generates the rotational symmetry of the $S^2$.

When $\frac{\Omega_{1E}}{\kappa_E}$ is a rational number $\frac{q}{p}$ for some coprime integers $p$ and $q$, the Killing vector field $\frac{\partial}{\partial \psi}$ generates a $U(1)$ action with finite orbits at infinity. 
It is not difficult to show that, for the $U(1)$ isometry subgroup generated by $\frac{\partial}{\partial\psi}$, this gravitational instanton has three nuts \cite{Gibbons:1979xm} located at the three turning points of its rod structure. For a preferred orientation of the manifold of this gravitational instanton, their corresponding nut-types are $(-q,p)$, $(p+q,q)$, and $(-p,-p-q)$ respectively. It is immediately clear from the results of \cite{Gibbons:1979xm} that the total nut charge of this gravitational instanton with respect to the Killing vector field $\frac{\partial}{\partial \psi}$ is zero, consistent with the AF asymptotic structure. The Euler number and Hirzebruch signature of this gravitational instanton are $\chi=3$ and $\tau=1$ respectively, which are the same as those of $\mathbb{C}P^2$. They can be computed by using either the formulae (4.6) and (4.7), or (4.1) and (4.2), in \cite{Gibbons:1979xm}. Note that since this gravitational instanton is AF, there is no contribution from the boundary terms to the topological invariants $\chi$ and $\tau$. Like the Taub-bolt instanton, this gravitational instanton does not admit a spin structure, and neither is its Riemann tensor nor its Weyl tensor self-dual or anti-self-dual.

The new AF gravitational instanton was constructed using the inverse-scattering method \cite{Belinski:2001,Pomeransky:2005sj}. We started with the Euclidean double-Schwarzschild black hole (in Weyl--Papapetrou coordinates) as the seed solution, and performed a four-soliton transformation following the techniques used in \cite{Pomeransky:2005sj}. In particular, four new Belinsky--Zakharov (BZ) parameters were introduced, one at each turning point of the seed. The BZ parameter introduced at the third turning point was then chosen in such a way that the third and fourth rods of the solution thus generated have identical directions, so that they join up to a single rod. In this way, the third turning point is actually eliminated, and the resultant solution has effectively three turning points. We then imposed the desired rod structure (Fig.~\ref{Figure_new_AF}) and brought the solution to the orientation such that $\ell_1=(0,1)$ and $g_{\psi\psi}=1$ at infinity. Finally, we tried to simplify the form of this solution as much as possible; the form (\ref{metric_new_AF}) was obtained after finding the suitable C-metric-like coordinates (\ref{Weyl_coordinates}). It turns out that in our construction, one parameter, say the BZ parameter introduced at the fourth turning point of the seed, is redundant. In hindsight, this parameter can be set to zero so that one can essentially use a three-soliton transformation to carry out the construction. 

One may wonder whether the new AF gravitational instanton is within the Euclidean version of the most general Ricci-flat type-D solution considered in \cite{Lapedes:1980qw}, since both their rod structures have three turning points. Despite intensive efforts, we have not been able to find such an identification. However, one may want to distinguish them from a deeper perspective, say by studying their algebraic types in the Euclidean regime.
It is also natural to wonder whether this gravitational instanton admits a Lorentzian section, like the Euclidean Kerr instanton. It is attractive to interpret such a section, if it exists, as a stationary double-Kerr configuration in the limit when the two event horizons touch each other. This configuration, however, must be singular by the classical black-hole uniqueness theorems.

The action of the new AF gravitational instanton is $I=\frac{\pi M}{\kappa_E}$, where $M=\frac{\varkappa^2}{2}(\gamma-\lambda^2)$ and $\kappa_E$ is given in (\ref{kappa_omega}). In calculating this action, we have used the flat space $S^1\times\mathbb{R}^3$ as the background, just as has been done for the Euclidean Kerr instanton \cite{Gibbons:1976ue}. Moreover, the action of this gravitational instanton turns out to be always greater than that of the Euclidean Kerr instanton, for fixed boundary conditions (identifications at infinity).
It deserves further study to see what role this gravitational instanton would play in the path integral of tunnelling amplitudes in Euclidean quantum gravity, and the contribution it would make to the partition function in the grand canonical ensemble. Nevertheless, we expect that this gravitational instanton should be included in the steepest-descents approximation of the path integral and the partition function.

In Kaluza--Klein theory, upon dimensional reduction, the Euclidean Kerr solution can describe a magnetic dipole \cite{Gross:1983hb,Sorkin:1983ns} separated by a string-like conical singularity. Similarly, upon dimensional reduction along the Killing vector field $\frac{\partial}{\partial \psi}$, the metric of our solution (\ref{metric_new_AF}) can describe a system of three collinearly positioned magnetic monopoles, say $p_1$, $p_2$ and $p_3$, located at the above-mentioned three nuts from left to right, and separated by string-like conical singularities. This system has zero total monopole charge but an in general non-vanishing dipole charge. Indeed, by direct calculation, the individual magnetic charges of the three monopoles sum up to zero, i.e., $p_1+p_2+p_3=0$. Moreover, it can be shown that $p_1$ and $p_3$ have the same sign, and opposite to that of $p_2$, so attractive forces dominate the magnetic interactions between these monopoles. Such a ``tripole'' system is balanced by the conical singularities between neighbouring monopoles, which always provide a strut force since we have $\Big|\frac{\Omega_{1,2E}}{\kappa_E}\Big|<1$. Here, we have assumed that $\phi$ in the reduced system is identified with period $2\pi$ to ensure asymptotic flatness. However, we remark that the metric (\ref{metric_new_AF}) is not the most general solution that describes a tripole system with zero total monopole charge. This is because in order to obtain our solution, we have imposed the regularity condition $\ell_3=\ell_1+\ell_2$, which is not necessary for a tripole system in Kaluza--Klein theory.

\begin{figure}[t]
\begin{center}
\includegraphics{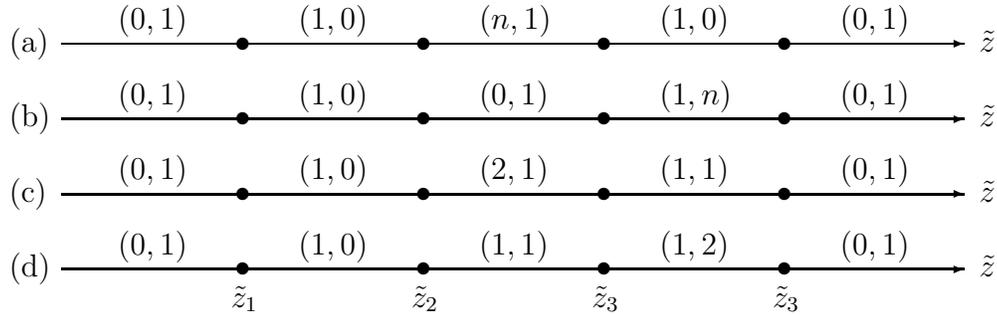}
\caption{The rod structures of possible new AF gravitational instantons with four turning points in standard orientation, for $n\in \mathbb{N}$. Note that (a) and (b) are the same if $n=0$.}
\label{Figure_new_AF_four_turning_point}
\end{center}
\end{figure}

Finally, it is of interest to see whether there are any other new AF gravitational instantons with an isometry group $\mathcal{T}$, and with $N>3$ turning points. In the case of $N=4$, the possible rod structures are shown in the four figures of Fig.~\ref{Figure_new_AF_four_turning_point}. Assuming the left and right semi-infinite rods in each figure join up to a single rod, the rod structures Fig.~\ref{Figure_new_AF_four_turning_point}(a,b), as well as Fig.~\ref{Figure_new_AF_four_turning_point}(c,d), are the same, and they correspond to compact gravitational instantons \cite{Chen:2010zu}. With this assumption, Fig.~\ref{Figure_new_AF_four_turning_point}(a,b) correspond to $S^2\times S^2$ when $n=0$, and to $\mathbb{C}P^2\,\sharp\, \overline{\mathbb{C}P^2}$ \cite{Page:1979zv} when $n=1$; Fig.~\ref{Figure_new_AF_four_turning_point}(a,b) with $n\geq 2$ and Fig.~\ref{Figure_new_AF_four_turning_point}(c,d) correspond to possible new compact gravitational instantons. Hence, the global topologies of the possible new AF gravitational instantons will be that of their corresponding compact ones with a circle $S^1$ removed appropriately. It remains an open problem to see whether they can be constructed using the inverse-scattering method, if they exist at all.

\bigbreak\bigskip\bigskip\centerline{{\bf Acknowledgement}}
\nobreak\noindent We would like to thank Gary Gibbons for some helpful comments and suggestions. This work was supported by the Academic Research Fund (WBS No.: R-144-000-277-112) from the National University of Singapore.

\bigskip\bigskip

{\renewcommand{\Large}{\normalsize}
}

\end{document}